\documentstyle[11pt,aas2pp4,psfig]{article}
 
\newcommand{\Ha}{H$\alpha$}
\newcommand{\Hb}{\ifmmode {\rm H}\beta \else H$\beta$ \fi}

\newcommand{\Ms}{M$_{\odot}$}
\newcommand{\kms}{km~s$^{-1}$}

\newcommand{\Zw}{I~Zw~18}
\newcommand{\hii}{\ion{H}{2}}
\newcommand{\heii}{HeII $\lambda$4686}
\newcommand{\Oiii}{[OIII] $\lambda$5007}
\newcommand{\civ}{CIV $\lambda$5808}
\newcommand{\erga}{erg~s$^{-1}$~cm$^{-2}$~\AA$^{-1}$}
\newcommand{\erg}{erg~s$^{-1}$~cm$^{-2}$}
\newcommand{\msun}{\ifmmode M_{\odot} \else M$_{\odot}$\fi}
\newcommand{\rsun}{\ifmmode R_{\odot} \else R_{$\odot}$\fi}
\newcommand{\lsun}{\ifmmode L_{\odot} \else L_{$\odot}$\fi}
\newcommand{\zsun}{\ifmmode Z_{\odot} \else Z_{$\odot}$\fi}
\newcommand{\izotov}{IFGGT}
\newcommand{\legrand}{LKRMW}
\newcommand{\HeII}{\ion{He}{2}}

\accepted{26 May 1988}

\slugcomment{Accepted for publication in ApJ}

\lefthead{de Mello et al.}
\righthead{I~Zw~18}

\begin{document}

\title{Searching for WR stars in \Zw\ --- The origin of \HeII\ emission
\footnote{Based on observations with the NASA/ESA {\it Hubble Space
Telescope}, obtained at the Space Telescope Science Institute, which
is operated by AURA, Inc., for NASA under contract NAS5-26555}
}

\author{Du\'\i lia F. de Mello}
\affil{Space Telescope Science Institute, 3700 San Martin Drive, Baltimore,
MD 21218 \\ E-mail: demello@stsci.edu}

\author{Daniel Schaerer\altaffilmark{2}}
\affil{Observatoire Midi-Pyr\'en\'ees, 14 Av. E. Belin F-31400, Toulouse, 
France \\ E-mail: schaerer@obs-mip.fr}

\author{Jennifer Heldmann}
\affil{Colgate University,
   Physics \& Astronomy, 13 Oak Drive, Hamilton, NY 13346\\
E-mail: jheldmann@center.colgate.edu}

\and
\author{Claus Leitherer}
\affil{Space Telescope Science Institute, 3700 San Martin Drive, Baltimore,
MD 21218 \\ E-mail: leitherer@stsci.edu}

\altaffiltext{2}{Space Telescope Science Institute, Baltimore, MD 21218}


\begin{abstract}

\Zw\ is the most metal poor star-forming galaxy known and is an ideal
laboratory to probe stellar evolution theory at low metallicities.  
Using archival HST WFPC2 imaging and FOS spectroscopy we were able to 
improve previous studies. We constructed a continuum free \heii\ map, 
which was used to identify Wolf-Rayet (WR) stars recently found
 by ground-based 
spectroscopy and to locate diffuse nebular emission.
Most of the \heii\ emission is associated with the NW stellar cluster, clearly
displaced
from the surrounding shell-like [\ion{O}{3}] and \Ha\ emission.
We found evidence for \HeII\ sources, compatible with 5--9 WNL stars
and/or compact nebular \heii\ emission, as well as residual diffuse
emission. Only one of them is outside the NW cluster. We have done 
an extensive comparison between our results and the 
recent ground-based data used by Izotov et al.\ (1997) and Legrand et al.\
(1997) to identify WN and WC stars in \Zw. 
The differences between the various data may be understood in terms of varying
slit locations, continuum fits, and contamination by nebular lines. 
We have calculated evolutionary tracks for massive stars and synthesis 
models at the appropriate metallicity ($Z$ $\sim 0.02$ Z$_{\odot}$). 
These single star models predict a mass limit $M_{\rm WR} \approx$ 
90 M$_{\odot}$ for WR stars to become WN and WC/WO.
For an instantaneous burst model with a Salpeter IMF extending
up to $M_{\rm up} \sim$ 120-150 \msun\ our model predictions
are in reasonable agreement with the observed equivalent widths. 
Our model is also able to fully reproduce the observed 
equivalent widths of nebular \heii\ emission due to the presence of WC/WO stars.
This quantitative agreement and the spatial correlation of nebular
\heii\ with the stellar cluster and the position of WR stars shown
from the ground-based spectra further supports the hypothesis
that WR stars are responsible for nebular \HeII\ emission in 
extra-galactic \hii\ regions. 

\end{abstract}



\medskip
	{\centerline\sl Accepted for publication in ApJ}

\twocolumn
\section{Introduction}
 
\Zw\ (= Mrk~116 = UGCA~166) is described in Zwicky's catalog of compact
galaxies (Zwicky 1971) as a pair of interconnected blue compact, presumably 
young galaxies. This otherwise unremarkable galaxy attracted immediate 
attention when Searle \& Sargent (1972) established its extremely low oxygen 
abundance of only 2\% the solar value. \Zw\ became one of the proto-types of 
the ``flashing galaxies'' (Searle, Sargent, \& Bagnuolo 1973), whose blue colors 
suggest a brief period of intense star formation. Subsequently this galaxy 
class was generally referred to as ``starburst galaxies'' (Weedman 1987), with 
\Zw\ belonging to the sub-category of blue compact dwarfs (Thuan 1991). 
A review of its
most pertinent parameters was given by Dufour et al. (1996): It has a radial 
velocity of 740~\kms, is at a distance of about 10~Mpc, and has an extinction
of $E(B-V) = 0.07$. 

\Zw's extreme metal deficiency was confirmed in several follow-up studies
(e.g., Kunth \& Sargent 1986; Campbell 1990; Pagel et al. 1992; Kunth et al.
1994; Stasi\'nska \& Leitherer 1996). Although strong efforts to detect other
galaxies with very low metal content have been undertaken (e.g. Terlevich,
Skillman, \& Terlevich 1995), \Zw\ is still the record holder. Its low 
metallicity is 
of particular importance for studies of the primordial helium abundance 
(Izotov, Thuan, \& Lipovetsky 1996; Olive, Skillman, \& Steigman 1997;
Skillman, Terlevich, \& Terlevich 1998), 
and for properties of chemically unevolved galaxies in general.
Despite the low oxygen abundance, however, there is clear evidence for
stellar nucleosynthetic processing from its carbon abundance so that 
\Zw\ is {\em not} a primordial galaxy (Garnett et al. 1997). The same
suggestion was made earlier by Thuan (1983) on the basis of surface
photometry of its underlying stellar population.

\Zw\ is an ideal laboratory to probe stellar evolution theory in a 
low-metallicity environment. The spatial resolution of WFPC2 on the
Hubble Space Telescope (HST) allows the detection of individual stars down
to the $\sim$10~\Ms\ range (Hunter \& Thronson 1995; Dufour et al. 1996). 
Analysis of deep WFPC2 images clearly demonstrates ongoing massive-star
formation. Additional, indirect evidence for a population of massive stars
comes from bubbles and outflows of the interstellar medium, indicative
of powerful hot-star winds and supernovae (Martin 1996).

Wolf-Rayet (WR) stars are the evolved, less massive descendants of previously 
massive O stars (Maeder \& Conti 1994). Stellar evolution theory predicts
few WR stars to form in a low-metallicity environment such as in
\Zw\ (e.g., Meynet 1995). Therefore recent reports (Izotov et al.\ 1997,  
hereafter \izotov; and Legrand et al.\ 1997, hereafter 
\legrand) of WR star detections in \Zw\  have far-reaching 
implications
for stellar evolution theory. 
Furthermore \Zw\ is probably the best studied object
among approximately 50 to 60 extra-galactic \hii\ regions showing nebular 
\heii\ emission (Campbell et al.\ 1986; Izotov, Thuan \& Lipovetsky 
1994, 1997)
whose origin is still poorly known and subject to recent debates.
Various mechanisms reaching from photoionization by hot massive stars, 
photoionization by X-ray binaries to shocks have been put forward to explain the high
excitation (Garnett et al.\ 1991). Understanding the origin of nebular \heii\ in 
low-metallicity galaxies has implications for our knowledge of the far-UV flux 
of galaxies and may prompt a re-examination of the importance of young galaxies 
to the ionization of QSO absorption line systems and the ionization of the 
intergalactic medium (cf. Garnett et al. 1991; Pettini et al. 1997).

WR stars can be recognized, e.g., via their broad
($\sim$100~\AA) emission bump around $\lambda$4686, which is generally 
a blend of H, \HeII\ and several metal lines. 
\Zw\ has been known to show {\em narrow}, nebular \heii\ emission but the 
{\em broad} WR emission bump was not known before
the findings of \izotov\ and \legrand.
The blending of several stellar and nebular emission lines around 4686~\AA\ and
the complex spatial morphology in ground-based data make it
challenging to disentangle stars and gas and to derive the WR content and the
nebular properties. This is evident in the apparently discrepant results of
\izotov\ and \legrand. The superior spatial 
resolution of HST may prove useful in resolving the discrepancy. 
 
In this paper we report on an HST archival project to study the spatial 
morphology of the highly ionized gas in \Zw, as observed primarily in \heii.
Our goals are to reconcile existing ground-based spectroscopic observations, 
to test stellar evolution models for WR stars in an extreme chemical
environment, and to elucidate the nature of nebular \HeII\ emission in
extragalactic \hii\ regions. The paper is organized as follows. In \S 2 and \S 3 we describe 
the data used in this work and their analysis. In \S 4
we discuss the recent spectroscopic detection of Wolf-Rayet stars in connection
with the HST imagery. The interpretation and a
comparison with Wolf-Rayet star models are presented in \S 5.  
Finally, \S 6 summarizes our conclusions.

\section{Data and reduction}
 
\subsection{Observation log}

We have retrieved HST archival data of \Zw\ obtained during Cycles~4 and 6. 
The data consist of both Wide Field/Planetary Camera 2 (WFPC2) images and 
Faint Object Spectrograph (FOS) spectra. The existence of such an extensive 
collection of data including images and spectra in several passbands make 
it possible to improve previous analysis, which were done using the available 
data at that time and which interpreted each data set individually
(e.g. Hunter \& Thronson 1995; Dufour et al. 1996). 
In Table~\ref{obs} we list the image root names (column~1), proposal
identification number (column~2), the epoch of observation (column~3), 
the configuration (column~4), and the optical element (column~5) for each set 
of data retrieved from the HST archive. In column~6 we give the spectral range
of each optical element. The entry is the full width at half maximum (FWHM)
for filters and the free spectral range for gratings. In the following 
subsections we describe several methods of data reduction which were tested 
using several different images and spectra.

\subsection{Pipeline processing}

All data were processed through the standard pipeline calibration.  Calibration
 of the FOS spectra in the Routine Science Data Processing (RSDP) system most 
notably includes detector background and scattered light subtraction, flatfield 
corrections, computation of wavelengths, and conversion from count rates to
fluxes.
The WFPC2 data were processed through the pipeline where bias, dark, 
and flatfielding corrections were performed and photometry keywords were 
calculated.

Since the expected flux levels are close to the background and the noise level 
of the detector, additional data reduction and analysis procedures on top of
the standard pipelines processing were required. These are described in the
following subsections. All steps were performed using standard tasks in 
IRAF/STSDAS. 

\subsection{Cosmic-ray and hot pixel removal}

Multiple images were simultaneously co-added and cosmic-rays removed 
using the IRAF task {\it crrej}.  We set the algorithm to estimate the 
initial value to `median' in order to eliminate the possibility of inadvertent 
rejection of true counts. Our study approaches the limits of HST detection, 
and high cosmic-ray flux values 
would bias the mean towards a higher number.  Likewise by setting the $\sigma$ 
threshold higher, we ensured the accuracy of the cosmic-ray rejection task.  
Surface plots of all supposed \heii\ emission were visually inspected 
for the characteristic profile of a cosmic-ray hit.
The probability that three cosmic-rays can hit the same pixel in 
three different images was tested by using dark images taken at the same 
epoch and with similar exposure time. The dark images were cleaned using
the same procedure as described above. We found that only 6 pixels remained 
contaminated by cosmic-rays (total number of pixels in each image = 6.4 $\times$ 10$^{6}$). 
Considering that this analysis will be critical when we discuss
the \heii\ sources in \S 3, it is important to notice that this number 
is even lower if we take into account that the area of the CCD where 
the helium sources are detected is much smaller (total number of 
pixels = 4.7 $\times$ 10$^{3}$). 
Therefore, our technique of eliminating cosmic-rays proved to be very 
efficient and adequate for this study.

All pixels with apparent 
\heii\ emission were also individually inspected and compared with the known 
coordinates of hot pixels on the chip to eliminate any erroneous detections.
 
\subsection{Background subtraction}

Background subtraction was also performed on each image.  Automatic background
subtraction through {\it crrej} which simply subtracts the mode of the image 
was not used because we determined that this method generally subtracted values 
much larger than the true background.  Instead, the background value was 
determined by computing 
the average of the counts calculated from the median of 13 boxes of 100 square 
pixels placed in relatively blank regions of each image.  This procedure was 
performed on all images and the corresponding background value subtracted.

\subsection{Flux calibration}

All images were then converted into absolute flux units.  Each image
 was multiplied by the appropriate {\it PHOTFLAM} value (\erga) which is 
defined to be the mean flux density that produces a count rate of 1 per second 
(DN) with the HST observing mode used in each observation.  The final flux 
calibrated image was created by dividing by the exposure time.

The reliability of this process was exhaustively confirmed through comparisons 
with several sources of spectral data.  Photometry was performed on the WFPC2 
images and compared with the flux values obtained from the FOS spectra.  
Measurements of spectra from proposal 6536 (y39a0304t) indicated line fluxes 
of $9.50\times 10^{-15}$~\erg\ and 
$7.59\times 10^{-15}$~\erg\ for \Ha\ and \Oiii\ ([OIII] hereafter), 
respectively.  These values refer to 
a circular entrance aperture with a diameter of 0.86$''$.  
We placed a synthetic aperture on the \Ha\ and [OIII] WFPC2 images and measured 
the encircled fluxes. Photometry was then performed on the \Ha\ and the
continuum subtracted [OIII] WFPC2 images generated from observations of 
proposal 5434. (See below for the adopted continuum subtraction technique.)
We measured $9.83\times 10^{-15}$~\erg\ for \Ha\ and 
$6.54\times 10^{-15}$~\erg\ for [OIII]. We consider these values in good 
agreement,
given the inherent uncertainties, most notably the unknown absolute pointing of 
HST.  Typical pointing uncertainties are $\pm 0.5''$. Photometry was also 
performed at offset positions approximately 1$''$ away from the assumed 
spectra locations to test the effect of HST pointing uncertainties.  
Flux differences on the order of 50\% were found.

Total flux levels of \Ha\ and [OIII] for the entire galaxy as measured from 
the flux calibrated WFPC2 images were likewise compared with published values
from ground-based spectra. In Table~\ref{fluxes} we compare our flux 
measurements with those of several different authors.
Our WFPC2 fluxes are for the entire galaxy. The published ground-based 
measurements have a spread of nearly an order of magnitude, with the extreme
values bracketing the WFPC2 values. Most likely, this reflects uncertainties
in the calibration of ground-based data, different filter 
systems, and the difficulty to measure faint extended emission on 
ground-based data. Our values are closer to Dufour \& Hester (1990) values 
which are based on imaging (see also Martin 1996). 
 
We are mostly concerned about the internal uncertainty of the WFPC2 photometry
since we combine different datasets to construct pure, continuum-free line
images. To address this concern, flux levels of multiple F555W images were 
compared. We used a WF3 and a PC image from proposal 
5434 and a PC image from proposal 5309.  Fluxes at discrete locations as well 
as the total flux of the main body of \Zw\ were compared in this filter.  
The total fluxes from these three images are $1.11\times 10^{-15}$~\erga, 
$1.14\times 10^{-15}$~\erga, and $1.10\times 10^{-15}$~\erga. 
We conclude from these cross-checks that the internal WFPC2 flux calibration 
is accurate to about 3\%.

\subsection{Continuum subtraction}

Pure emission-line images were produced by subtracting the underlying 
continuum from the appropriate narrow-band image.  A pure oxygen map was 
created by subtracting a scaled continuum image compatible with the 
wavelength range of the F502N filter. We also created an \Ha\ map.  
The \Ha\ emission was taken as the flux in
the F658N filter as the \Ha\ line is redshifted to $\sim$6580~\AA, 
which is within the bandwidth of the F658N filter (FWHM~=~28.5~\AA). 
A very conservative upper limit of 10\% for the continuum contribution was
estimated from FOS spectra. [NII] $\lambda$6548/83 is outside the passband
of the F658N filter. Hence we use the flux calibrated F658N image 
as an \Ha\ map without any further correction.

A continuum-free spatial map of \heii\ was created using the F336W image since 
this image contains no significant line emission.
Based on the FOS spectra of proposal 6536, the ratio of the continuum fluxes 
of the F469N and F336W filters at their effective wavelengths is 0.35.  The 
F336W image was scaled by the appropriate factor and 
then subtracted from the F469N for a pure \heii\ map. 
It is important to notice that the FOS spectra were centered on a small region 
(diameter = 0.86$''$) on the southwest side of the NW region of \Zw\ and 
not necessarily represent the entire NW region. In order to check this we 
have compared FOS fluxes with ground based fluxes (F. Legrand 1997, 
private communication). We found that the FOS spectra are quite 
representative of the entire NW region, i.e. larger scales. However,
in smaller scales the same cannot be assured.
To put the following
discussion into perspective, we note that the pure \heii\ flux is about 60\%
of the continuum flux within the F469N filter. 

The emission-line images used for the analysis were obtained as described in the
previous paragraphs. The adopted continuum subtraction technique was the
result of extensive tests using different methods. Ideally, one would like
to use a narrow-band image in the adjacent, line-free continuum for the
continuum subtraction. Given HST's filter set, such data do not exist. Errors
in the continuum subtraction process are introduced by using (i) continuum 
images which are contaminated by lines and (ii) continuum images whose 
effective wavelength is very different from that of the line image. In the
following we discuss our experiments to evaluate the trade-offs between (i)
and (ii). Since our main interest is in the \HeII\ map, we restrict our discussion
to the continuum correction of the F469N filter.

Continuum maps were created by correcting the F555W images for emission lines. 
As determined through equivalent width measurements of FOS spectra from 
proposal 6536 (y2f90402t), the emission lines in this filter contribute about
40\% to the total flux of the F555W filter, and so this fraction was 
subtracted to generate the continuum map.  These corresponding 
continuum images were then subtracted from the F469N image to create a \HeII\ 
map. However, the large correction required to account for the line 
contamination in the F555W filter and its probable spatial dependence
introduce a large uncertainty. The resulting \HeII\ maps turned out to be
unreliable. 

Other experiments involved various combinations of the F450W and F439W 
images together with F336W and F555W to inter/extrapolate the continuum level
at 4686~\AA. Again, we noticed the uncertain line contamination correction
for the continuum filters. This was tested by determining the location
of the continuum using F336W (no significant contamination), F439W and
F450W (some contamination), and F555W (strong contamination). Our tests
showed that by far the more important error source in creating a \HeII\ map is
the line contamination in the continuum filters and their spatial variation.
In contrast, the stellar continuum slope is not strongly variable. This is
immediately obvious: stars contributing to the optical/near-ultraviolet light
have spectral types O to early A. Their $(B-V)$ colors are in the narrow
range of --0.3 to +0.1 so that the continuum slope is fairly constant and has 
little spatial dependence.

We conclude from these simulations that our \heii\ map is well-suited to 
detect any excess \HeII\ emission but that its absolute flux calibration must
be considered with care. We adopt a conservative error of 30\%. Next we 
discuss the H$\alpha$, [OIII], and \heii\ maps.

\section{Data Analysis}

\begin{figure*}[htp]
\centerline{\psfig{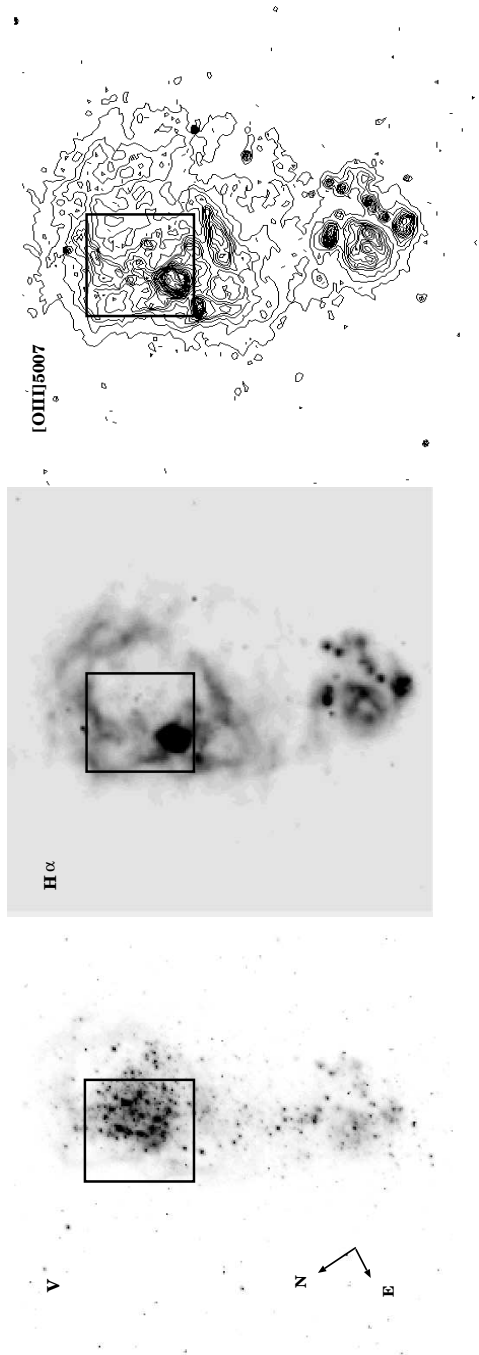}}
\caption{WFPC2 V (F555W),  H$\alpha$ (F658N) and [OIII] $\lambda$5007 (F502N) 
image of \Zw\. The V image was taken with the PC, the H$\alpha$ and [OIII] 
images were taken with the WF3 and transformed to PC resolution. The rectangle 
delineates the area where the continuum-free helium sources were detected. 
Darker regions in the 
H$\alpha$ image represent the strongest fluxes. [OIII] contours are from 
1$\times$10$^{-19}$ to 2.6$\times$10$^{-18}$ \erga\ (15 levels). Size of
rectangle is $3'' \times 3.3''$ (157 pc $\times$ 172~pc). Orientation is 
given on the bottom-left corner of the V image.
\label{f1}} 
\end{figure*}

In Fig.~\ref{f1} we show the WFPC2 V (F555W),  H$\alpha$ (F658N) and 
(F502N) 
images of \Zw\.  The total H$\alpha$ 
and [OIII] fluxes are given in Table~\ref{fluxes}, which was described in \S 2.5. 
The H$\alpha$ and [OIII] maps have a very similar morphology with coincident peaks 
of emission and similar filamentary structure. 
The ionized gas distribution is displayed in forms of `bubbles' probably 
driven by stellar winds and supernovae. The nebular peaks of H$\alpha$ and 
[OIII] are spatially offset from the U, V and I continuum. We refer to 
Hunter \& Thronson (1995) and Martin (1996) for a more detailed analysis 
of the ionized gas and its relationship to the stars.
 
\begin{figure*}[htp]
\centerline{\psfig{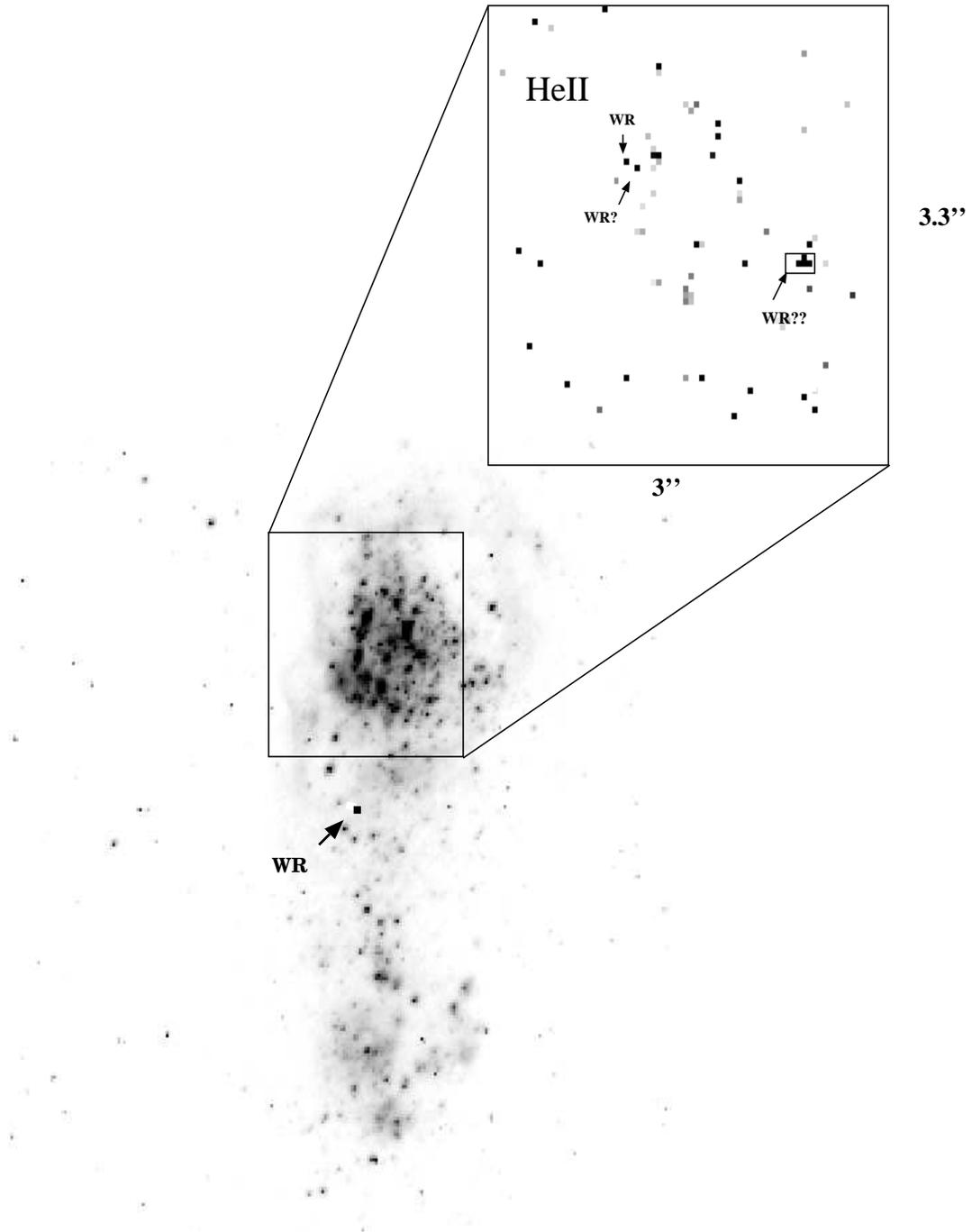}}
\caption{ WFPC2 V (F555W) image of \Zw\. The rectangle delineates 
the area in the NW region where
the continuum-free helium sources were detected. The darkest pixels in the 
helium
map are above the 3$\sigma$ level. ``WR'' and ``WR?'' identify helium 
sources with fluxes equivalent to 3 WNL and 0.7 WNL stars, respectively.  
``WR??'' identifies sources equivalent to 0.4, 0.7, 1.2, and 2 WNL stars.
``WR'' outside the NW region identifies a helium source equivalent to 
1.5 WNL stars.
\label{f2}}
\end{figure*}

The spatial map of continuum-free \heii\  was carefully analyzed and 25 
individual pixels of \HeII\ emission greater than 3$\sigma$ above the 
`background' 
level have been identified within an area of 9.9 arcsec$^{2}$ (62 pixels $\ge$ 2$\sigma$). 
This area is centered 
on the NW component of the galaxy and is presented in Fig.~\ref{f2} together with the V 
image. 
The darkest pixels in Fig.~\ref{f2} have fluxes higher than 3$\sigma$. We have 
verified that these \HeII\ sources are neither contaminated by cosmic-rays 
nor by hot pixels (see \S 2.3 for more details on cosmic-ray removal).
As a final check we identified each \heii\ emission in the three individual
images used in {\it crrej}. Since the probability that three cosmic-rays hit
the same pixel in three different images is very low (see \S 2.3) this 
procedure guarantees that each emission detected
in the three images are not cosmic-rays, i.e. pixels that were not detected 
in each of the three images were considered contaminated by cosmic-ray.

From the present data we have no direct means of distinguishing nebular 
from stellar \heii\ emission. 
Stars of the types Of, Ofpe/WN, WNL, and WNE show \heii\ emission formed
in the expanding stellar wind. A fraction of the broad blended emission
of \ion{C}{3}/\ion{C}{4} $\lambda\lambda$ 4640-4650 and \heii\ in WC and 
WO stars may also be detected in the F469N filter. 
The average {\em stellar} \HeII\  line luminosity of WR stars of the WN sequence is 
$L_{\rm 4686}({\rm WNE})=(5.2 \pm 2.7) \times 10^{35}$ erg~s$^{-1}$, and
$L_{\rm 4686}({\rm WNL})=(1.6 \pm 1.5) \times 10^{36}$ erg~s$^{-1}$ 
(see Schaerer \& Vacca 1998 for line luminosities of other types).
Placing a WNL star at the distance of \Zw\ (10.8 Mpc) yields a typical line flux of 
$1.15\times 10^{-16}$~\erg. The \HeII\ filter (F469N) has a width of 24.9 {\AA} and 
therefore the average line flux of a WNL star detected with F469N would 
be expected to be of the order of 
$4.6\times 10^{-18}$~\erga \ (average line flux over the FWHM of the filter). 
However, we should take into account the fact that Galactic WN stars have \heii\ 
13 {\AA}$<$FWHM$<$63 {\AA} and WN stars in the SMC have 17 {\AA}$<$FWHM$<$33 {\AA} 
(Conti \& Massey 1989; Conti et al. 1989). Considering that I~Zw~18's stars are more 
similar to SMC's stars than to the ones in the Milky Way, we conclude that the average 
line flux detected with the F469N (FWHM=24.9 {\AA}) contains most of the true emission. 
 
The total \HeII\ flux coming from helium sources ($\ge$3$\sigma$) in the 
region presented in Fig.~\ref{f2} is 
$3.97\times 10^{-17}$~\erga, 
equivalent to $\sim$ 9 WNL stars \footnote{The flux value is not corrected for 
extinction. Using the values of Hunter \& Thronson (1995) for A$_{\lambda}$ (0.21, 0.13, and 0.08 
for F336W, 
F555W, and F814W, respectively) and assuming that the Milky Way extinction curve is adequate 
for dereddening, we find A$_{4686}$ = 0.18.}. 
However, this flux is distributed over several regions. For instance, 12 
individual 
pixels are above the 5$\sigma$ level and have fluxes higher than 
$1.0\times 10^{-18}$~\erga. We have identified 6 of these sources as 
possible WR stars ($\ge$10$\sigma$). The pixels marked with WR and ``WR?'' 
in Fig.~\ref{f2} have a flux of $1.3\times 10^{-17}$~\erga \ ($\sim$ 3 WNL stars) and 
$3.4\times 10^{-18}$~\erga \ ($\sim$0.7 WNL star), respectively.
The pixels marked with ``WR??'' have fluxes of
1.7, 3.0, 5.3 and 9.2 $\times 10^{-18}$~\erga, which is equivalent to 
0.4, 0.7, 1.2 and 2 WNL stars (4.3 WNL in total), 
respectively. However, only one of these pixels 
(0.4 WNL) is classified as an unambiguous detection. The other three
pixels are present in two of the images but are seen as faint detections 
in one of the images. Therefore, we cannot exclude the possibility of contamination 
by cosmic-rays (see \S 2.3 for more details on cosmic-ray removal).

Although the fluxes from the individual \HeII\ sources found are compatible
with fluxes of WNL stars, we cannot exclude sources of compact nebular \heii\ emission
or a combination of stellar and nebular emission.
Placed at the distance of \Zw\ and observed with the F469N filter the 3 
nebular \heii\ sources DR1, SMC N76, and LMC N44C studied by Garnett et al.\ (1991) 
have fluxes between $1.6\times 10^{-17}$ (N76) and $1.8\times 10^{-18}$~\erga\ (N44C),
comparable to 0.4-3.6 WNL stars.
Their angular diameters range from $\sim$ 0.4$''$ (DR1) to 0.03 $''$ (N44C) 
at the distance of \Zw.
Therefore, our \HeII\ detections are also compatible with nebular sources 
similar to Garnett's objects provided that they have compactness similar to
N44C. 

\begin{figure}[htp]
\centerline{\psfig{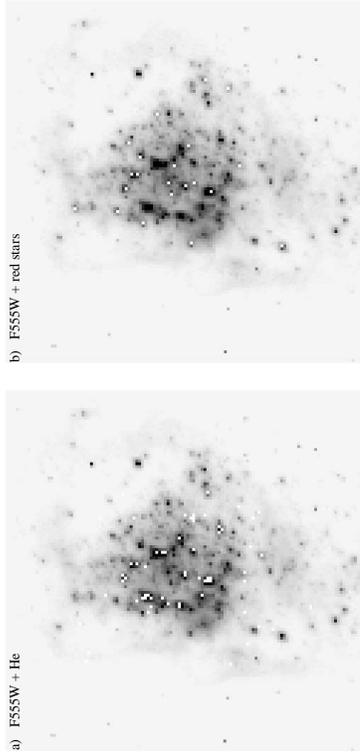}}
\caption{WFPC2 V (F555W) image of \Zw\ a) The white rectangles 
correspond to the Helium sources from F469N image. b) The white rectangles 
correspond to the red stellar objects identified by Hunter \& Thronson (1995).
\label{f3}}
\end{figure}

The integrated line flux within the 9.9 arcsec$^{2}$ region including the helium sources 
is found to be $1.67\times 10^{-16}$~\erga, \ i.e.\ $4.16\times 10^{-15}$~\erg.
For comparison, ground-based observations of the narrow \heii\ emission in the
NW component of \Zw\ yield somewhat lower values between $\sim 5.0 \times 10^{-16}$ and 
$1.3 \times 10^{-15}$~\erg\ (Pagel et al.\ 1992; Skillmann \& Kennicutt 1993, 
\izotov; Izotov \& Thuan 1998), which is likely explained by the smaller area
covered by these observations.
In our images the difference between the total flux and the flux from the helium 
sources is 1.1$\times 10^{-16}$~\erga. This residual emission is spread over the 
9.9 arcsec$^{2}$ region and is probably nebular. We have also checked for the
possibility of contamination by the red stellar objects found by 
Hunter \& Thronson (1995) since these objects may contribute flux at 4690 \AA\ 
but not in F336W. In Fig.~\ref{f3} we show the helium sources and all the 
red stars brighter than F555W=25.0 and redder than F555W--F814W=0.5 (see Hunter 
\& Thronson 1995 for details on magnitude calculation). We find that 5 of 
these red stars coincide with the position where helium sources were identified. However,
these helium sources are very faint with fluxes of $\sim$ $5-9\times 10^{-19}$~\erg.
The total helium emission from these sources contributes less than 6\% to
the total flux in the entire region.

In order to compare the features of different maps with the \heii\ map, Wide Field 
(WFC) images of H$\alpha$ (u2f90205t and u2f90206t) and [OIII] (u2f90203 and 
u2f90204) 
were rebinned into Planetary Camera (PC) resolution. We used the task {\it 
magnify} 
in IRAF to perform the image interpolation. We have used 5 V images 
(u2cg0201t, u2cg0202t, u2cg0203t, u2f90104t and u2f90105t) taken with the 
PC and WF3 in order to 
check the validity of this technique. We conclude that the transformed images 
preserve enough information on the general morphology of 
the maps for a qualitative inspection. 
However, the transformed image loses information during the image interpolation 
and should be used with caution when comparing detailed features.

The helium peaks are located in the NW cluster, whereas the [\ion{O}{3}] and 
H$\alpha$ emission has its maximum in a shell-like structure surrounding the cluster.
Measured in small circular areas around the possible WR stars one finds
$I(4686)/I({\rm H}\alpha) \la$ 0.3-0.1; a fairly constant value of 
$I(5007)/I({\rm H}\alpha) \approx 1$ is obtained over the entire region of the cluster. 
However, these values should be taken with caution since they were obtained using
the WF images transformed to PC resolution.

Several stars are identified in the area of the \HeII\ peaks, including the second
brightest V star in the list of Hunter \& Thronson (1995). This star is only 
4 pixels 
(1 pixel=0.046$''$) to the west of the WR stars and according to those authors it 
has magnitudes F555W=22.43, F336W=21.67 (see Hunter \& Thronson 1995 for details
on their photometry). 

We have also investigated the presence of helium sources outside the NW cluster. 
We found that one source has flux $\sim$1.5 WNL. It is located in between 
the NW and the SE condensations. In Table~\ref{numbers} we summarize the 
helium sources identified in I~Zw~18. 
 
From the spatial maps presented in this section we conclude that \heii\ emission
is clearly associated with the stellar cluster and spatially offset from the 
maximum nebular H$\alpha$ and [OIII] emission. The present data alone do not 
allow us to distinguish between nebular and stellar \heii\ emission.  

\section{Comparison with ground-based spectroscopy}

Two very recent papers have reported different numbers of WR stars in
\Zw\ using ground-based spectroscopy.  LKRMW estimated from spectra taken 
with the 3.6m CFH Telescope that 1-2 WC4 or WC5 and no WN stars are present 
in \Zw. In contrast, IFGGT using the Multi Mirror Telescope found 17$\pm$4 WNL 
and 5$\pm$2 
WC4 stars.  Both spectra unambiguously show broad blue and red emission bumps 
centered at 
$\sim$4645-4686 {\AA} and at 5808 {\AA}, respectively, but differ in the measured 
line
fluxes and the structure of the blue emission bump, as summarized below.
There may be several reasons why their results differ. The main reason
may be the choice of a different slit position and orientation.
The slits were centered approximately on the same region (central knot of NW 
\hii\ 
region). However, the position angle used by LKRMW  is $+45^{\circ}$ 
compared to $-41^{\circ}$ by IFGGT.
Therefore, their spectra cover partially different regions of the galaxy. 
 
For the following discussion and for comparisons with theoretical predictions of the 
massive star population in
\Zw\ (see \S 5) we complement both sets of measurements of LKRMW and IFGGT with 
equivalent width measurements
kindly provided by F. Legrand (1997, private communication), and with additional \Hb\ 
equivalent widths
from the literature. The data are summarized in Table~\ref{lines}. Before proceeding
we give the following cautionary 
remarks: (i) LKRMW detect a broad component centered at $\sim$4645 \AA\ (cf.\ ``broad 
4645'' in Table~\ref{lines}) and no broad emission at $\lambda \approx$ 4686 \AA. 
A small change in the fit of the continuum level may, however, 
accommodate the existence of a broad component.  
(ii) IFGGT detect a broad ``WR bump'' extending over $\lambda \approx$ 4619 -- 4740 {\AA}
(``broad blue bump'' in Table~\ref{lines}). As mentioned by these authors
 this feature can generally be composed of several broad WR emission 
 lines as well as nebular lines. The presence of [FeIII] $\lambda$4658, 
 [ArIV]+HeI $\lambda\lambda$4711,4713, and [ArIV] $\lambda$4740 can be 
suspected in their spectrum (the latter two lines are present in LKRMW).
When present, these nebular emission lines typically have line 
intensities of 0.01 to 0.03 \Hb\ in total, as can be seen from the data of 
Izotov, Thuan \& Lipovetsky (1994, 1997). 
The contribution due to WR stars may therefore be overestimated by up to 
a factor of two. (iii) Both the intensity of \civ/\Hb\ and the \civ\ 
line flux integrated over the respective regions differ approximately by a 
factor of two between LKRMW and IFGGT. This is very likely explained by the 
complex spatial distribution of the 
\Hb\ emission as seen from the HST maps and the different adopted 
slit positions/orientations.
(iv) The observations of the nebular \heii/\Hb intensity ($\sim$0.040) 
of LKRMW and IFGGT are in agreement but are slightly larger than previous measurements 
(\heii/\Hb $\approx$ 0.032; e.g.\ Pagel et al.\ 1992; Skillman \& Kennicutt 
1993). 
(v) The \Hb\ equivalent widths measured for the NW HII region differ 
considerably 
between various authors. Values between 56 and 127 \AA\ are found by 
Pagel et al.\ (1992), Skillman \& Kennicutt (1993), LKRMW and IFGGT.

Comparing the ground-based results of LKRMW and IFGGT with our results 
(see Table~\ref{numbers} for a summary our final numbers) it is 
possible that LKRMW have underestimated the number of WN stars while 
\izotov\
et al. have overestimated it.
The HST helium filter used, F469N, covers the region of 4683-4707 {\AA} and 
therefore 
the helium emission line which is redshifted to 4698 {\AA} is well centered in 
the filter 
passband. There is no contamination by the lines that are blended in the 
4645 {\AA} bump; 
the two closest lines, [FeIII] $\lambda$4658, [ArIV] $\lambda$4711 are redshifted to 
4669 {\AA} and 4722 {\AA}, 
respectively, lying outside the helium filter. Therefore the helium emission 
detected 
can be either stellar or nebular in origin. However, we cannot exclude the presence 
of WC stars since there are no HST 
data available for the region of the 5808 {\AA} bump. 

The absence of WN stars was inferred by LKRMW from the absence of broad 4686 {\AA}
emission underlying the narrow 4686 {\AA}. However, a small change 
in their fit of continuum level would provide enough flux to account for a few WN stars.
It is also possible that no WN stars were detected because none were in the region 
where their spectra were taken; their slit width is 1.52$''$ and 
could be centered just below the region where the stars are (as discussed at the
end of this section). 

The number of WNL stars derived by IFGGT is most likely 
overestimated
for the following reasons: (i) The contribution of WC stars to the blue WR bump 
($\lambda \approx 4686$ {\AA})
has not been accounted for, and (ii) the broad \heii\ bump may be contaminated by 
other nebular
emission lines as discussed above.
Indeed the observed flux ratio $f(4686)/f(5808) \approx 2$ is typical for WC4-5 
stars
(cf.\ Smith et al.\ 1990; Schaerer \& Vacca 1998) and the contribution of 
CIII/IV $\lambda$4650 from the WC stars to the 4686 {\AA} bump should be subtracted to 
determine
the remaining luminosity from WNL stars.
Using the recent compilation of WR line luminosities from Schaerer \& Vacca 
(1998)
we obtain: $N_{\rm WC4} \sim 4$, and $N_{\rm WNL}=4$. Note that the main 
difference between the value $N_{\rm WNL}=17 \pm 4$ obtained by IFGGT is due 
to (i). If (ii) were also taken into account, the inferred number of WNL stars may be even lower.
In any case it appears that the differences between the estimates of the number 
of WR 
stars derived by IFGGT  and LKRMW  are  not significant if uncertainties are taken
into account. Future observations with higher disperion and signal-to-noise will 
be necessary to address these issues with more accuracy.
In absolute terms, the number of WNL stars in particular may be more uncertain
given the relatively large spread of \heii\ line luminosities of individual WNL 
stars (Schaerer \& Vacca 1998, their Fig.\ 1).

\begin{figure}[htp]
\centerline{\psfig{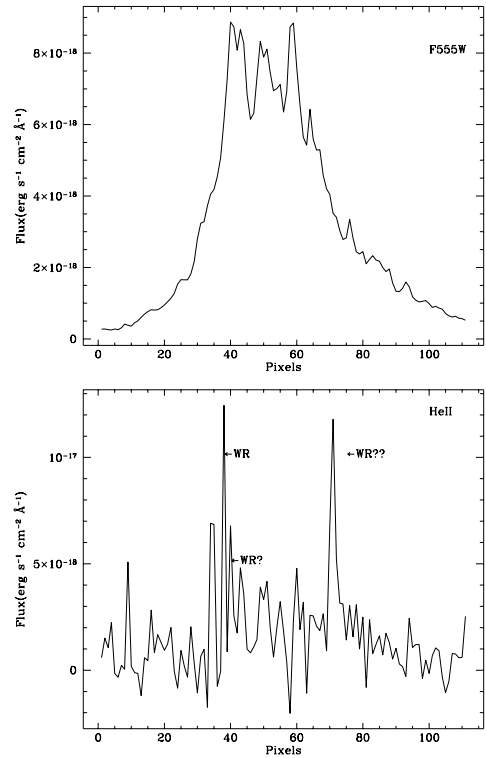}}
\caption{Summed flux within a region of 1.5$''$ along the NW region of 
V (F555W) and HeII (F469N) (pixel size is 0.046$''$).
\label{f4}}
\end{figure}

 
\legrand\ and Izotov \& Thuan (1998) have analyzed the spatial location of several
emission features in \Zw. 
\legrand\ have found that the bumps at 4645 {\AA} and 5820 {\AA}
peak between 1$''$ and 2$''$ SW from the central cluster. They have also found 
a spatial correlation
between WC features and nebular \heii. In order to reproduce their results we
have added (IRAF task {\it improject}) the flux of the HST images within 
1.5$''$ (their slit width). In Fig.~\ref{f4} we present this result for the helium and 
F555W images (pixel size is 0.046$''$). 
Three helium peaks are easily identified; two of them are the ``WR'' and ``WR?'' and
the third peak is the one marked with ``WR??'' in Fig.~\ref{f2}. Comparing the 
helium peaks with the F555W peaks we conclude that the ``WR??'' peak is probably the one 
claimed by LKRMW as the peak located SW of the central cluster. Due to the lack of resolution 
present in LKRMW spectra, the central cluster is probably centered in between the 
three peaks in the F555W (around pixel 50). ``WR??'' is 1$''$ away from
pixel 50, which agrees with the distance found by LKRMW. However, as we 
discussed in \S 3 there is the possibility of cosmic-ray contamination at the ``WR??'' position. 
Our results suggest that LKRMW's slit location probably missed the ``WR'' and ``WR?'' peaks, 
otherwise one bump should have been detected on the NE of 
the central cluster. The two peaks, ``WR'' and ``WR?'', cannot be resolved 
with the ground-based telescopes since they are only 0.1$''$ apart situated on the NE of 
the central cluster. The low spatial resolution and uncertainties in pointing
of the ground-based spectra limit the comparison with HST data. 

The shift between the hydrogen emission and the continuum found by \legrand\ agrees
well with the HST images. Our spatial maps are also compatible with the 
spatial distribution found by Izotov \& Thuan (1998). The results of \izotov\
do not provide information about the position of the WR features.

\section{Model comparison and interpretation}
 
In order to compare all the data with appropriate model predictions of
massive star population at the low metallicity of \Zw\, we have first calculated 
a set of new stellar evolution tracks and then performed evolutionary synthesis
calculations using these tracks. Preliminary results have been discussed in
Schaerer (1997). The ingredients of the stellar evolution models 
and a discussion of the results from these calculations are presented 
in the next subsections.

\subsection{Stellar evolution models}
\label{s_stellar}

Evolutionary tracks for stars between 25 and 150 M$_{\odot}$\ 
($M_{\rm ini}=25, 40, 60, 85, 120, 150)$ at the metallicity 
$Z=0.0004 \approx 1/50$ Z$_{\odot}$ have been calculated with the Geneva stellar 
evolution code 
adopting the same ingredients as Meynet et al.\ (1994). 
These calculations extend the previously available Geneva sets ($0.001 \le Z \le 
0.1$)
down to the metallicity of \Zw. The adopted mass-loss rate and its metallicity 
dependence
are the main factors determining the evolution of the most massive stars.
As shown by Maeder \& Meynet (1994) the high mass-loss rates adopted in the 
models of Meynet et al.\ (1994) reproduces  many observational properties of 
individual WR stars and O star populations at different metallicities, although
direct measurements of WR mass-loss rates at Z$_{\odot}$ suggest lower values (Leitherer
et al. 1997). We have therefore adopted the same mass-loss prescription and 
scaling with metallicity. In addition we have also computed evolutionary tracks 
using mass-loss rates for OB
stars determined from the recent results of Lamers \& Cassinelli (1996), which
relies on the observed wind-momentum-radius relation by Kudritzki et al.\ 
(1995).
This prescription also explicitly includes a metallicity dependence, based on 
observations of O stars in the Galaxy, the LMC, and the SMC.

Qualitatively, the evolutionary tracks at $Z=0.0004$ reproduce the properties found 
earlier (cf.\ Meynet et al.\ 1994). Due to the diminished mass 
loss only the
most massive stars are predicted to evolve to WR stars. The (initial) mass limit 
of single 
WR stars is found to be $M_{\rm WR} \approx$ 90~M$_{\odot}$, compared to 61 M$_{\odot}$ at 
$Z=0.001$
(Maeder \& Meynet 1994).  Both the 120 and 150 M$_{\odot}$\
models evolve through the WN phase prior to becoming WC/WO stars shortly before 
the 
end of He-burning.
At the entry to this phase the ratio of the surface abundances is (C+O)/He $>$ 1 by 
number since the core is revealed in the late stage of He-burning. Following Smith \& 
Maeder (1991)
these stars would be classified as WO. 
For stars with initial masses $\ga$ 80 M$_{\odot}$\ the use of the mass-loss 
prescription of 
Lamers \& Cassinelli (1996) leads to essentially identical mass loss during the
main-sequence (MS) evolution as with the adopted high mass-loss rates. This reflects 
the weak metallicity dependence of mass loss at high luminosity (see Lamers 
\&
Cassinelli 1996).
The subsequent evolution, the WR mass limit, and other relevant properties are 
therefore
nearly identical to the high mass loss case, although they are of course still subject 
to
uncertainties in post-MS mass-loss rates.
In summary, our models predict WNL stars, a short WNE phase, and highly evolved 
WC/WO stars for single stars with initial masses $M_{\rm ini} \ga$ 90 M$_{\odot}$\ at 
the 
metallicity of \Zw.

\subsection{Evolutionary synthesis models -- comparisons with observations}
\label{s_synthesis}

Subsequently we have calculated evolutionary synthesis models using the stellar 
tracks
described above. The calculations are done with the models of Schaerer \& Vacca 
(1998)
where a detailed description of the input physics can be found.
The most relevant features are the use of spherically expanding non-LTE atmospheres 
to describe the ionizing fluxes and the synthesis of WR emission features in the 
optical spectrum based on the
recent compilation of line luminosities by Schaerer \& Vacca (1998).
We consider a power-law IMF (e.g.\ Salpeter) with a variable upper 
mass cut-off 
$M_{\rm up}$, and instantaneous star-formation at time $t=0$ (``instantaneous 
burst'').
If the evolution of the observed WR stars in \Zw\ is described by the new single 
star models, $M_{\rm up} > M_{\rm WR} \approx$ 90 M$_{\odot}$ \ is required. Such a 
high value is very well compatible with observations of massive stars in the 
Magellanic Clouds (Puls et al. 1996).
The choice of the lower mass cut-off does not affect our results.  

\begin{figure}[htp]
\centerline{\psfig{figure=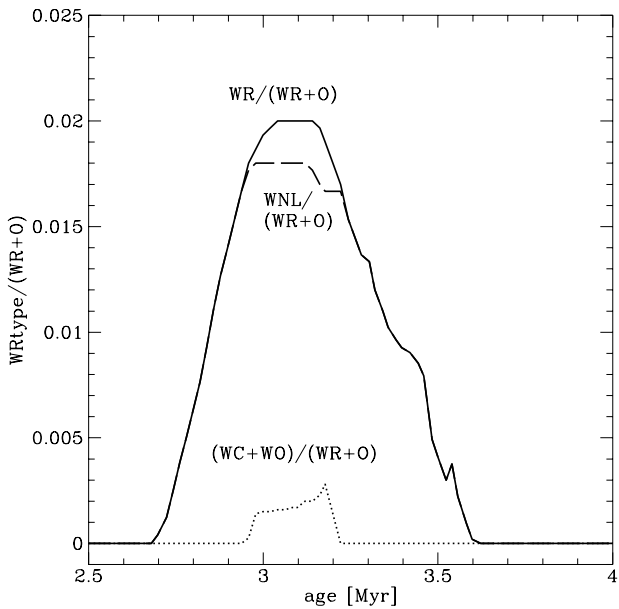,width=8.8cm}}
\caption{Predicted number ratio WR type/(WR+O) as function of 
time at $Z=$0.0004 for the instantaneous burst model with a Salpeter IMF. 
Solid: total WR/(WR+O), dashed: WNL/(WR+O), dotted: (WC+WO)/(WR+O)
\label{f5}}
\end{figure}

\begin{figure}[htp]
\centerline{\psfig{figure=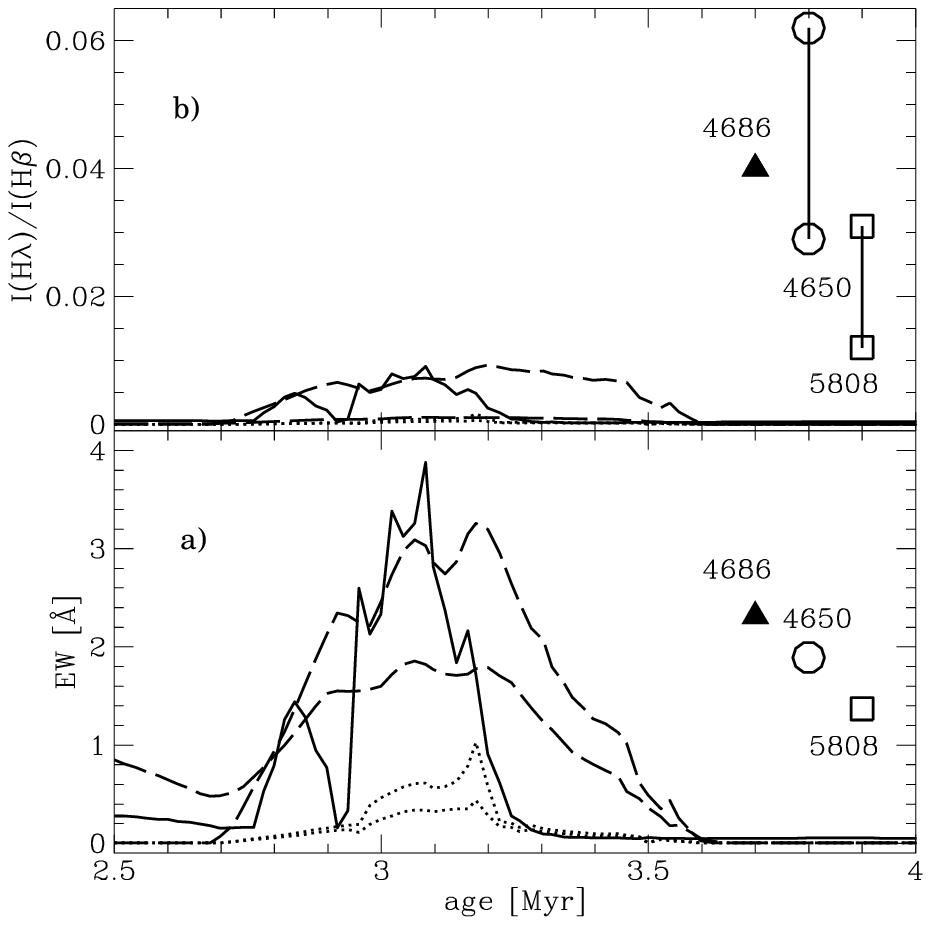,width=8.8cm}}
\caption{Comparison of observed and predicted values WR and nebular
\protect\heii\ features for \Zw. Observed values (cf.\ Table \ref{lines}) for
nebular \heii\ (solid triangles), broad 4645 or 4686 emission (denoted 4650; 
circles), and \civ\
(squares) are shown. Differing measurements from IFGGT and LKRMW for 4650 and 5808
are connected. Predicted values as a function of time are given for nebular \heii\
(solid line), the 4650 bump (dashed lines), and \civ\ (dotted lines) for an
instantaneous burst model with a Salpeter IMF.
The two values for the dashed and dotted lines correspond to different
prescriptions for the WR line luminosities.
{\em lower panel} a): equivalent widths, 
{\em Upper panel} b): line intensities relative to \Hb.
\label{f6}}
\end{figure}

In Fig.~\ref{f5} we show the predicted WR/(WR+O), WNL/(WR+O) and WC+WO/(WR+O) number 
ratios
for $M_{\rm up}=$150 M$_{\odot}$ as a function of the age of the burst. WR stars 
appear after
$\sim$2.5 Myr and their presence lasts approximately 1 Myr. The maximum WR/(WR+O) 
ratio
is $\sim$0.02. As expected from models at $Z=0.001$, WN stars dominate the WR 
population;
the predicted WC/WN ratio is $\la 0.15$.
The predicted WR lines (line intensities with respect to \Hb\ and equivalent 
widths)
and the prediction for the nebular \heii\ emission are plotted in Fig.~\ref{f6}
as a function of time.
Also shown are the observed quantities from IFGGT and LKRMW 
 summarized in Table~\ref{lines}.
It is evident that the spatial complexity of the gas and 
the displacement between stars and gas cause strong variations in the stellar 
and nebular emission strengths.
This also explains the large variations of $W(\Hb)$ shown in Table~\ref{lines}.
On the other hand, the WR features and, interestingly, also nebular \heii, follow 
fairly 
closely the spatial distribution of the continuum emission (see IFGGT, LKMRW, 
and the HST images). Therefore, we prefer to use the equivalent 
widths of the WR features as the physically most meaningful quantity for 
quantitative comparisons. Taking the largest $W(\Hb)$ 
value (127 \AA) from the literature, a lower limit of $\sim$3 Myr can be estimated 
for the age of the NW region from our instantaneous burst models.

Fig.~\ref{f6}b shows that the observed {\it line intensities} exceed all 
predicted
values by a large factor. This is expected due to the complex morphology of the
interstellar gas.
In contrast, the predicted {\it equivalent widths} fall in the range of 
the observed values from LKRMW.
We have plotted the prediction for the 
CIII/IV $\lambda$4650 + \HeII\ $\lambda$4686 bump using the calibration from Smith 
(1991) (curves with larger values in Fig.~\ref{f6}a), and the sum of NIII $\lambda$4640 and 
CIII/IV $\lambda$4650 from the compilation of Schaerer \& Vacca (1998) (curves with 
lower values in Fig.~\ref{f6}a). These predictions bracket the observed value 
of LKRMW and could even marginally explain
the larger flux measured by IFGGT.
The predicted $W(5808)$ is generally somewhat lower than the 
observations. Using the line luminosities of Smith (1991) for WC/WO stars we predict
$W \la$ 1 \AA; lower values are obtained using the values from the
compilation of Schaerer \& Vacca, which accounts for differences between
WC and WO subtypes. In view of the observational uncertainties (see discussion in \S 4)
we conclude that the present single star models are able
to reproduce the WR features in \Zw\ discussed so far with an instantaneous burst
and a Salpeter IMF extending to $M_{\rm up} \approx$ 120--150 M$_{\odot}$.
 
Interestingly, the predicted equivalent width of nebular \heii\ reproduces
the observed value very well (Fig ~\ref{f6}a). 
The total nebular \HeII\ emission in \Zw\ is consistent
with the derived number of WC stars if one assumes that their spectrum
has the same hardness (i.e., identical ratio of He$^{++}$/H$^{+}$ continuum 
photons) as the well studied WO star DR1 (Garnett et al.\ 1991, Kingsburgh 
et al.\ 1995), and that the Lyman continuum luminosity corresponds to that 
of an average WR star at low $Z$ (Schaerer \& Vacca 1998, their Fig.\ 6). 
Both facts support the hypothesis
of Schaerer (1996) that WC/WO stars are responsible for nebular \heii\ 
features observed in extragalactic HII regions.
This is in contrast to the result of IFGGT  and LKRMW, 
who conclude, based only on the comparison of the observed \HeII/\Hb\ 
line intensity, that the nebular \HeII\ emission in \Zw\ cannot be 
explained quantitatively by WC/WO stars. The apparent contradiction is
simply explained by the different spatial extension of the \HeII\
and \Hb\ emission seen in the HST imagery (cf.\ \S 4). The spectroscopic
\HeII/\Hb\ values in IFGGT and LKRMW probably overestimate the true
nebular L(\HeII)/L(\Hb) ratio because \Hb\ is emitted over a larger region than 
the \HeII. The need for such a geometric correction was pointed out by 
Garnett et al. (1991).
An additional argument from \Zw\ in favor of the existence of an 
intimate link between WC/WO stars and nebular \HeII\ emission 
is the spatial correlation between the WC features and \heii\ found 
by LKRMW.

\subsection{WC and/or WO stars in \Zw}
\label{s_wo}

As mentioned in \S \ref{s_stellar} the stellar models predict
a very high C+O and a low helium abundance in the WC/WO phase, which would likely lead to a
WO instead of WC classification of these stars.
Are WC and/or WO stars present in \Zw?
A classification of the WC and WO subtypes observed in \Zw\ requires more than 
the mere detection of \civ\ and the 4650 bump (Smith et al.\ 1990; Crowther et 
al.\ 1998).
However, as discussed by LKRMW, the observed FWHM of these lines 
is compatible with WC4 or WC5 types (cf.\ Smith et al.).
The width of \civ\ given by both IFGGT  and LKRMW  and
the structure of the 4650 bump observed by the former authors are
also compatible with those of the WO3-4 star DR1 (Kingsburgh et al. 1995; Crowther 
et al.\ 1998).
If the entire \civ\ emission were attributed to WO3-4 stars, one would expect
I(OV $\lambda$5590)/I(CIV) $\approx 0.15$ (Schaerer \& Vacca 1998),
too weak to be detected in the current observations. A detection of 
the broad OIV $\lambda$3400, OVI $\lambda$3811, or 
OV $\lambda$5590 lines is required for an unambiguous detection
of WO stars. At this time the presence of WO stars, expected in low metallicity
environments, cannot be excluded in \Zw,
and the classification of the observed WC stars is presently not certain.

If the fraction of evolved WR stars with 
(C+O)/He $>$1 (``WO'') is low among the observed WC/WO stars, 
the present evolutionary scenario may not
apply and other channels leading to an earlier appearance of He-burning
products on the stellar surface might be required.
Two (non-exclusive) processes may be invoked: higher mass loss on the main sequence 
and/or the WNL phase or additional mixing leading to upward transport
of processed elements.
Given the uncertainties of the mass-loss rates of very massive stars
and their metallicity dependence (cf.\ Heap et al.\ 1994; Schaerer 1998)
the evolutionary scenarios of single stars are still uncertain.
Extreme mass loss may also occur through Roche-lobe overflow in a 
binary system. Indeed as shown by Schaerer (1998) Case A binaries 
(Roche-lobe filled during MS phase) may be an important channel to
form WR stars from high mass stars ($M_{\rm initial} \ga$ 40 M$_{\odot}$)
in low $Z$ environments. 
Rotational induced mixing can lead to an earlier entry in the WR
phase (e.g.\ Meynet 1997; Meynet \& Maeder 1997).
New observations of both individual massive stars or integrated 
populations in low metallicity environments will provide crucial
information to guide our understanding of massive stars in extreme environments.




\section{Summary}

We have used an extensive collection of archival HST data to study the stellar 
content and gaseous distribution in \Zw.
Continuum subtraction and cosmic-ray removal were found to be critical for 
the search for faint sources. 
Using WFPC2 imaging and FOS spectra we were able to improve previous 
studies and construct a continuum free \heii\ map, which was used to 
search for stellar sources (WR stars) recently found by ground-based 
spectroscopy and to locate diffuse nebular emission.
The \heii\ emission is associated with the NW stellar cluster, clearly
displaced
from the surrounding shell-like [\ion{O}{3}] and \Ha\ emission.
We found evidence for \HeII\ sources, compatible with 5--9 WNL stars 
and/or compact nebular \heii\ emission, as well as residual diffuse
emission. Only one of these sources is not located in NW cluster but 
in the region in between the NW and SE condensation. 

We have done an extensive comparison between our results and the
recent ground-based data of \izotov\ and \legrand\
which allowed them to identify WN and WC stars in \Zw. The
differences between the various data may be understood in terms of varying
slit locations, continuum fits and contamination by nebular lines. 
Ground-based spectra taken with better resolution and larger telescopes
will allow one to probe the massive star content more accurately, to provide more stringent
constraints on evolution models, and to study the interaction between
the stars and their surrounding ISM in more depth.
In order to avoid the difficulties
in correlating spectra and image location, Fabry-Perot interferometry or 
other imaging spectrocopy
would be the technique of choice. 

We have calculated evolutionary tracks for massive stars and synthesis 
models at the appropriate metallicity ($Z$ $\sim 0.02$ Z$_{\odot}$). 
These single star models predict a mass limit $M_{\rm WR} \approx$ 
90M$_{\odot}$ for WR stars which may become WN and WC/WO.
For an instantaneous burst model with a Salpeter IMF extending
up to $M_{\rm up} \sim$ 120-150 \msun\ our model predictions
are in reasonable agreement with the observed equivalent widths. 
The WR population in \Zw\ has properties consistent with those predicted
by single-star evolution models for this metallicity. The WR/O ratio forms an
extension of the trend observed in WR galaxies (Maeder \& Conti 1994) towards
very low $Z$. 
Our model is also able to fully reproduce the observed 
EW of nebular \heii\ emission due to the presence of WC/WO stars.
This quantitative agreement and the spatial correlation of nebular
\heii\ with the stellar cluster and the position of WR stars 
further supports the hypothesis by
Schaerer (1996) that WR stars are responsible for nebular \HeII\ emission in 
extra-galactic \hii\ regions.

\acknowledgments

We thank the anonymous referee for valuable comments which helped improving 
our paper. We are grateful to Stefano Casertano, Bill Vacca and Rosa 
Gonzalez-Delgado for helpful discussions.
D.S. is grateful to Yuri Izotov and Fran\c{c}ois Legrand for helpful 
discussions and for providing us with additional information from their spectra.
Deidre Hunter kindly sent us her original images for comparison purposes. 
J.H. acknowledges support from STScI summer student program.
D.S. acknowledges support from the Swiss National Foundation for Scientific
Research and partial support from the Director's Discretionary Research Fund 
of the STScI.

\clearpage


\begin{table}
\scriptsize
\caption{HST archival data of \Zw \label{obs}}
\begin{center}
\begin{tabular}{c c c c c c}
\hline
\hline
\multicolumn{1}{c}{Root Name} &
\multicolumn{1}{c}{Proposal ID} &
\multicolumn{1}{c}{Epoch}&
\multicolumn{1}{c}{Configuration} &
\multicolumn{1}{c}{Optical Element} &
\multicolumn{1}{c}{$\Delta \lambda$ (\AA)}\\
\hline
y2f90403t &5434&27 Oct 1994 &FOS/RD &G190H&722\\
y2f90404t &5434&27 Oct 1994 &FOS/RD &G190H&722\\
y2f90402t &5434&27 Oct 1994 &FOS/RD &G570H&2249\\
u2cg0101t &5309&29 Oct 1994 & PC    &F336W&371\\
u2cg0102t &5309&29 Oct 1994 & PC    &F336W&371\\
u2cg0103t &5309&30 Oct 1994 & PC    &F336W&371\\
u2cg0301t &5309&30 Oct 1994 & PC    &F814W&1758\\
u2cg0302t &5309&30 Oct 1994 & PC    &F814W&1758\\
u2cg0303t &5309&30 Oct 1994 & PC    &F814W&1758\\
u2cg0401t &5309&31 Oct 1994 & PC    &F469N&25\\
u2cg0402t &5309&31 Oct 1994 & PC    &F469N&25\\
u2cg0403t &5309&31 Oct 1994 & PC    &F469N&25\\
u2cg0201t &5309&31 Oct 1994 & PC    &F555W&1223\\
u2cg0202t &5309&31 Oct 1994 & PC    &F555W&1223\\
u2cg0203t &5309&31 Oct 1994 & PC    &F555W&1223\\
u2f90203t &5434&02 Nov 1994 &WF3    &F502N&27\\
u2f90204t &5434&02 Nov 1994 &WF3    &F502N&27\\
u2f90205t &5434&02 Nov 1994 &WF3    &F658N&29\\
u2f90201t &5434&02 Nov 1994 &WF3    &F702W&1481\\
u2f90202t &5434&02 Nov 1994 &WF3    &F702W&1481\\
u2f90101t &5434&03 Nov 1994 &WF3    &F702W&1481\\
u2f90102t &5434&03 Nov 1994 &WF3    &F450W&925\\
u2f90103t &5434&03 Nov 1994 &WF3    &F450W&925\\
u2f90104t &5434&03 Nov 1994 &WF3    &F555W&1223\\
u2f90105t &5434&03 Nov 1994 &WF3    &F555W&1223\\
u2cg0501t &5309&03 Nov 1994 & PC    &F656N&22\\
u2cg0502t &5309&03 Nov 1994 & PC    &F656N&22\\
u2cg0503t &5309&03 Nov 1994 & PC    &F656N&22\\
u2f90206t &5434&03 Nov 1994 &WF3    &F658N&29\\
u2f90303t &5434&01 Mar 1995 & PC    &F439W&464\\
u2f90304t &5434&01 Mar 1995 & PC    &F439W&464\\
u2f90301t &5434&01 Mar 1995 & PC    &F555W&1223\\
u2f90302t &5434&01 Mar 1995 & PC    &F555W&1223\\
u2f90305t &5434&01 Mar 1995 & PC    &F675W&889\\
u2f90306t &5434&01 Mar 1995 & PC    &F675W&889\\
y39a0305t &6536&08 Jun 1996 &FOS/RD &G190H&722\\
y39a0306t &6536&08 Jun 1996 &FOS/RD &G190H&722\\
y39a0303t &6536&08 Jun 1996 &FOS/RD &G400H&1546\\
y39a0304t &6536&08 Jun 1996 &FOS/RD &G570H&2249\\
\hline
\end{tabular}\end{center}
\end{table}

\clearpage

\begin{table}
\caption{Flux comparisons \label{fluxes}}
\begin{center}
\begin{tabular}{l c c}
\hline
\hline
\multicolumn{1}{c}{Authors\tablenotemark{a}} &
\multicolumn{1}{c}{\Ha\ Flux} &
\multicolumn{1}{c}{\Oiii\ Flux}\\
\multicolumn{1}{c}{} &
\multicolumn{1}{c}{(\erg)} &
\multicolumn{1}{c}{(\erg)}\\
\hline
This work  & $3.3\times 10^{-13}$  & $2.0\times 10^{-13}$\\
DH         & $4.2\times 10^{-13}$  & $2.4\times 10^{-13}$\\
DKF        & $3.9\times 10^{-13}$\ & $2.8\times 10^{-13}$\\
IFGGT      & $1.2\times 10^{-13}$  & $8.3\times 10^{-14}$\\
PSTE       & $1.8\times 10^{-13}$  & $4.6\times 10^{-14}$\\
SK\tablenotemark{b}&1.1 $\times 10^{-13}$&$7.0\times 10^{-14}$\\
\hline
\end{tabular}
\tablenotetext{a}{DH: Dufour \& Hester (1990); DKF: Davidson et al. (1989); IFGGT: Izotov et al. (1997); 
PSTE: Pagel et al. (1992); SK: Skillman \& Kennicutt (1993)}
\tablenotetext{b}{Note that only SK fluxes are reddening corrected}
\end{center}
\end{table}


\begin{table*}
\caption{WR stars identified in \Zw. \label{numbers}}
\begin{center}
\begin{tabular}{lccl}
\hline
\hline

\multicolumn{1}{c}{Identification} & 
\multicolumn{1}{c}{Position\tablenotemark{a}}&
\multicolumn{1}{c}{Fluxes\tablenotemark{b}}& 
\multicolumn{1}{c}{Comments}\\
\multicolumn{1}{c}{}&
\multicolumn{1}{c}{}&
\multicolumn{1}{c}{(WNL)}&
\multicolumn{1}{c}{}\\
\hline 
WR  & 403, 465 & 3.0 & in NW cluster \\  
WR? & 405, 464 & 0.7 & in NW cluster \\ 
WR?? & 437, 449 & 0.4 & SW of the NW cluster\\
WR?? & 436, 450 & 0.7 & cosmic-ray contamination? \\
WR?? & 435, 449 & 1.2 & cosmic-ray contamination? \\
WR?? & 436, 449 & 2.0 & cosmic-ray contamination? \\
WR   & 406, 407 & 1.5 & between NW and SE clusters\\
\\
1--2 WC&     &\legrand\\
17$\pm$4 WNL&    &\izotov\\
5$\pm$2 WC4&     &\izotov\\
\hline
\end{tabular}
\tablenotetext{a}{x,y coordinates of each pixel in the F469N image}
\tablenotetext{b}{The average line flux of 1 WNL star observed with the 
F469N filter at the distance of I~Zw~18 is 4.6$\times 10^{-18}$\erga}
\end{center}
\end{table*}


\begin{table*}
\caption{Observed broad and nebular \heii\ in \Zw. \label{lines}}
\begin{center}
\begin{tabular}{lccl}
\hline
\hline

\multicolumn{1}{c}{Feature} & 
\multicolumn{1}{c}{$I_\lambda/I({\rm H_\beta})$}& 
\multicolumn{1}{c}{W$_\lambda$} & 
\multicolumn{1}{c}{Authors\tablenotemark{a}} \\
\multicolumn{1}{c}{}&
\multicolumn{1}{c}{}&
\multicolumn{1}{c}{(\AA)}&
\multicolumn{1}{c}{}\\
\hline 
Broad 4645	& 0.029	& 1.8 & LKRMW, L97 \\  
Broad blue bump ($\sim$4686) & 0.062 & & IFGGT \\
Broad 5808 	& 0.012	-- 0.031 & 1.3 & LKRMW, IFGGT, L97 \\
\\
Nebular \heii\ 	& 0.04	& 2.2	& LKRMW, IFGGT, L97 \\
\Hb\		& 1.00	& 56 -- 127 & IFGGT, PSTE\\
\hline
\end{tabular}
\tablenotetext{a}{LKRMW: Legrand et al. 1997, L97: Legrand  
(1997; private communication), IFGGT: Izotov et al. 1997, PSTE: Pagel et al. (1992)}
\end{center}
\end{table*}
  
\clearpage











%
%

\clearpage

\clearpage 

\begin{references}

\reference{} Campbell, A. 1990, \apj, 362, 100

\reference{} Campbell, A., Terlevich, R., \& Melnick, J. 1986, \mnras, 223, 811



\reference{} Conti, P.S., \& Massey, P. 1989, \apj, 337, 251

\reference{} Conti, P.S., Massey, P., \& Garmany, C.D. 1989, \apj, 341, 113

\reference{} Crowther, P.A., De Marco, O., Barlow, M.J. 1998, \mnras,
in press 

\reference{} Dufour, R.J., Garnett, D.R., Skillman, E.D. \& Shields, G. 1996,
in From Stars to Galaxies, ASP Conf. Series, Vol. 98, eds. C. Leitherer, U. 
Fritze-v. Alvensleben, and J. Huchra, 358 

\reference{} Dufour, R.J., \& Hester, J.J. 1990, ApJ, 350, 149


\reference{g91} Garnett, D.R., Kennicutt R.C., Chu, Y.-H., \& Skillman,
E.D. 1991, \apj, 373, 458

\reference{} Garnett, D. R., Skillman, E. D., Dufour, R. J., \& Shields, G. A. 1997, 
   ApJ, 481, 174 

\reference{} Heap S.R., Ebbets, D., Malumuth, E.M., Maran, S.P., de
Koter, A., \&
Hubeny, I. 1994, \aap, 435, L39

\reference{} Hunter, D., \& Thronson, H. 1995, \apj, 452, 238

\reference{} Izotov, Y.I., Foltz, C.B, Green, R.F., Guseva, N.G., \&
Thuan, T.X. 1997, ApJ, 487, L37 (IFGGT)

\reference{} Izotov, Y.I., \& Thuan, T.X. 1998, \apj, 497, 227

\reference{} Izotov, Y.I., Thuan, T.X., \& Lipovetsky, V.A. 1994, \apj,
435, 647

\reference{}  -------- 1997, ApJS,
108, 



\reference{} Kingsburgh, R.L., Barlow, M.J., \& Storey, P.J. 1995, \aap,
295, 75

\reference{} Kudritzki, R.P., Lennon, D.J., \& Puls, J. 1995, in Science
with the VLT,
eds. J.R. Walsh, I.J. Danzigers, 246

\reference{} Kunth, D., Lequeux, J. Sargent, W.L.W., \& Viallefond, F. 1994, 
\aap, 282, 709

\reference{} Kunth, D., \& Sargent, W.L.W. 1986, \apj, 300, 496

\reference{} Lamers, H.J.G.L.M., \& Cassinelli, J.P. 1996, in From Stars
to Galaxies,
ASP Conf. Series, Vol. 98, eds. C. Leitherer, U. Fritze-v. Alvensleben,
and J. Huchra, 162

\reference{} Legrand, F., Kunth, D., Roy, J.-R., Mas-Hesse, J.M., \&
Walsh, J.R. 1997,
	\aap, 326, L17 (LKRMW)

\reference{} Leitherer, C., Chapman, J.M., \& Koribalski, B. 1997, \apj, 481, 898

\reference{} Maeder, A., \& Conti, P. 1994, ARA\&A, 32, 227

\reference{} Maeder, A., \& Meynet, G. 1994, \aap, 287, 803

\reference{} Martin, C. 1996, \apj, 465, 680


\reference{} Meynet, G. 1995, \aap, 298, 767

\reference{} ------ 1997, Boulder Munich Workshop II: in Properties of Hot Luminous 
Stars, Vol. 131, ed. I. Howarth, ASP Conf.~Series, 96


\reference{} Meynet, G., Maeder, A., Schaller, G., Schaerer, D., \&
Charbonnel, C.
	1994, \aaps, 103, 97


\reference{} Olive, K.A., Skillman, E.D., \& Steigman, G. 1997, \apj, 489, 1006

\reference{} Pagel, B.E.J., Simonson, E.A., Terlevich, R.J., \& Edmunds,
M.G. 1992,
	\mnras, 255, 325


\reference{} Pettini, M., Smith, L., King, D.L, \& Hunstead, R.W. 1997, \apj, 
486, 665  

\reference{} Puls, J., Kudritzki, R.P., Herrero, A., Pauldrach, A.W.A.,
Haser, S.M., Lennon, D.J.,
Gabler, R., Voels, S.A., Vilchez, J.M., Wachter, S., \& Feldmaier, A.,
1996, \aap, 305, 171 
 
\reference{heii} Schaerer, D. 1996, \apjl, 467, L17

\reference{} ------ 1997, in Dwarf Galaxies: Probes for Galaxy
Formation and Evolution,
ed. J. Andersen, Highlights of Astronomy, in press

\reference{heii} ------ 1998, \aap, in preparation

\reference{} Schaerer, D., \& Vacca, W.D. 1998, \apj, 497, 618

\reference{} Searle, L., \& Sargent, W. L. W. 1972, ApJ, 173, 25

\reference{} Searle, L., Sargent, W.L.W., \& Bagnuolo, W.G. 1973, \apj, 179, 427

\reference{} Skillman, E.D., \& Kennicutt, R.C.Jr. 1993, \apj, 411, 655

\reference{} Skillman, E.D., Terlevich, E., \& Terlevich, R. 1998
in Primordial Nuclei and their Galatic Evolution, eds. N. Prantzos,
M. Tosi, \& R. van Steiger, Kluwer, in press



\reference{s91} Smith, L.F. 1991, in Wolf-Rayet Stars and Interrelations
with Other
        Stars in Galaxies, IAU Symp.~143, eds. K. A. van der Hucht \& B.
Hidayat,
        (Dordrecht: Kluwer), p.~601

\reference{} Smith, L.F., \& Maeder, A. 1991, \aap, 241, 77

\reference{} Smith, L.F., Shara, M.M., \& Moffat, A.F.J. 1990, \apj,
348, 471

\reference{} Stasi\'nska, G., \& Leitherer, C. 1996, \apjs 107, 661

\reference{} Terlevich, E., Skillman,E.D., Terlevich, R., 1995,
in The Interplay between Massive Star Formation, the ISM, and Galaxy
Evolution, eds. D. Kunth, B. Guidernoni, M. Heydari-Malayeri,
T.X. Thuan, Editions Fronti\`eres, Gif-sur-Yvette, 395

\reference{} Thuan, T.X. 1983, \apj, 268, 667

\reference{} ------ 1991, in Massive Stars in Starbursts, eds.
   C. Leitherer, N. Walborn, T. Heckman, \& C. Norman
   (Cambridge: CUP), 183

\reference{} Weedman, D. W. 1987, in Star Formation in Galaxies, ed. C. J. 
   Lonsdale (Washington: NASA), 351

\reference{} Zwicky, F. 1971, in Catalogue of selected compact galaxies 
and of post-eruptive galaxies Publ. F. Zwicky, G\"umligen (BE), 
Switzerland-388


\end{references}
\end{document}